%% file: main.tex
\documentclass[sigconf]{acmart}

\setcopyright{none}
\renewcommand\footnotetextcopyrightpermission[1]{}
\makeatletter
\renewcommand\@formatdoi[1]{\ignorespaces}
\makeatother

\AtBeginDocument{%
  \providecommand\BibTeX{{%
    \normalfont B\kern-0.5em{\scshape i\kern-0.25em b}\kern-0.8em\TeX}}}

\acmConference[Arxiv 2024]{Arxiv}{2024}{Online}

\usepackage{amsmath,amsfonts}
\usepackage{graphicx}
\usepackage{textcomp}
\usepackage{color}
\usepackage{hyperref}
\usepackage{adjustbox}
\usepackage{makecell}
\usepackage{multirow}
\usepackage{threeparttable}
\usepackage{enumitem}
\usepackage{booktabs}
\usepackage{bbding}
\usepackage{colortbl}
\usepackage{xcolor}
\usepackage{setspace}
\usepackage{pifont}
\usepackage{caption}
\usepackage{subcaption}
\usepackage{url}
\usepackage{listings}
\usepackage[linesnumbered,ruled,vlined]{algorithm2e}
\usepackage[noend]{algpseudocode}
\usepackage[most]{tcolorbox}
\usepackage{pifont}
\usepackage{indentfirst}
\usepackage{tabularx}

\newcommand{\codellama}{Code Llama}
\newcommand{\starcoder}{StarCoder}

\newcommand{\ourmethod}{ProCC}

\newcommand{\PromptRetrieval}{\textit{prompt-based multi-retriever system}}
\newcommand{\PromptBandit}{\textit{adaptive retrieval selection algorithm}}
\newcommand{\CodeContext}{lexical semantics}
\newcommand{\ExactMatch}{Exact Match}

\definecolor{codegreen}{rgb}{0,0.6,0}
\definecolor{codegray}{rgb}{0.5,0.5,0.5}
\definecolor{codepurple}{rgb}{0.58,0,0.82}
\definecolor{backcolour}{rgb}{0.95,0.95,0.92}

\lstdefinestyle{mystyle}{
    backgroundcolor=\color{backcolour},   
    commentstyle=\color{codegreen},
    keywordstyle=\color{magenta},
    numberstyle=\tiny\color{codegray},
    stringstyle=\color{codepurple},
    basicstyle=\ttfamily\footnotesize,
    breakatwhitespace=false,         
    breaklines=true,                 
    captionpos=b,                    
    keepspaces=true,                 
    numbers=left,                    
    numbersep=5pt,                  
    showspaces=false,                
    showstringspaces=false,
    showtabs=false,                  
    tabsize=2
}

\newtcolorbox{mybox}{
  enhanced,
  colback=gray!20,
  boxrule=1pt,
  arc=1mm,
  outer arc=1mm,
  boxsep=2mm,
  colframe=black,
  halign title=center,
  top=1mm,
  bottom=1mm,
  left=1mm,
  right=1mm,
  width=0.95\linewidth,
  before=\vspace{10pt},
  after=\vspace{10pt}
}

\begin{document}


\title{Prompt-based Code Completion via Multi-Retrieval Augmented Generation}

\author{Hanzhuo Tan}
\affiliation{%
  \institution{Southern University of Science and Technology and The Hong Kong Polytechnic University}
  \country{China}
}
\email{hanzhuo.tan@connect.polyu.hk}

\author{Qi Luo}
\affiliation{%
  \institution{Southern University of Science and Technology}
  \country{China}
}
\email{12232440@mail.sustech.edu.cn}

\author{Ling Jiang}
\affiliation{%
  \institution{Southern University of Science and Technology}
  \country{China}
}
\email{11711906@mail.sustech.edu.cn}

\author{Zizheng Zhan}
\affiliation{%
  \institution{Kwai Inc.}
  \country{China}
}
\email{zhanzizheng@kuaishou.com}

\author{Jing Li}
\affiliation{%
  \institution{The Hong Kong Polytechnic University}
  \country{China}
}
\email{jing1li@comp.polyu.edu.hk}

\author{Haotian Zhang}
\affiliation{%
  \institution{Kwai Inc.}
  \country{China}
}
\email{zhanghaotian@kuaishou.com}

\author{Yuqun Zhang}
\affiliation{%
  \institution{Southern University of Science and Technology}
  \country{China}
}
\email{zhangyq@sustech.edu.cn}

\input{sections/0_abstract}

\maketitle

\input{sections/1_introduction}
\input{sections/2_background}
\input{sections/3_approach}

\input{sections/4_evaluation}
\input{sections/5_results}
\input{sections/6_threats}
\input{sections/7_related_work}
\input{sections/8_conclusion}
\newpage

\bibliographystyle{ACM-Reference-Format}
\bibliography{ref}

\appendix

\end{document}

%% file: sections/0_abstract.tex
\begin{abstract}
Automated code completion, aiming at generating subsequent tokens from unfinished code, has been significantly benefited from recent progress in pre-trained Large Language Models (LLMs). However, these models often suffer from coherence issues and hallucinations when dealing with complex code logic or extrapolating beyond their training data. Existing Retrieval Augmented Generation (RAG) techniques partially address these issues by retrieving relevant code with a separate encoding model where the retrieved snippet serves as contextual reference for code completion. However, their retrieval scope is subject to a singular perspective defined by the encoding model, which largely overlooks the complexity and diversity inherent in code semantics.
To address this limitation, we propose \ourmethod{}, a code completion framework leveraging prompt engineering and the contextual multi-armed bandits algorithm to flexibly incorporate and adapt to multiple perspectives of code. \ourmethod{} first employs a \PromptRetrieval{} which crafts prompt templates to elicit LLM knowledge to understand code semantics with multiple retrieval perspectives. Then, it adopts the \PromptBandit{} to incorporate code similarity into the decision-making process to determine the most suitable retrieval perspective for the LLM to complete the code. Experimental results demonstrate that \ourmethod{} outperforms state-of-the-art code completion technique by 8.6\% on our collected open-source benchmark suite and 10.1\% on the private-domain benchmark suite collected from a billion-user e-commerce company in terms of \ExactMatch{}. \ourmethod{} also allows augmenting fine-tuned techniques in a plug-and-play manner, yielding 5.6\% improvement over our studied fine-tuned model. 

\end{abstract}

%% file: sections/1_introduction.tex
\section{Introduction}\label{sec:intro}
\cite{btc,slade,nova,scaling-law,anghabench,humaneval}Automated code completion aims at generating subsequent code tokens based on the ongoing incomplete code segments~\cite{convention_complete1, convention_complete2, convention_complete3, code_complete_Ngram, code_complete_attention, code_complete_rnn, code_complete_LSTM, lu2022reacc, zhang2023repocoder, knm}. Typically, it can significantly enhance the efficiency of software developers and reduce operating costs for corporations~\cite{savekeystroke,code_expect}.
Therefore, automated code completion has become widely incorporated into Integrated Development Environments (IDEs). For example, Microsoft's Pyright~\cite{pyright}, a static type-checking tool, powers the auto-completion feature for Python in VScode. Conventional automated code completion techniques generally rely on program analysis to generate syntax-conforming completions ~\cite{convention_complete1, convention_complete2, convention_complete3}. However, they are often argued to be limited in capturing code semantics to generate line-level completions for real-world development~\cite{code_expect}. To alleviate such issues, many existing techniques~\cite{code_complete_Ngram, code_complete_attention, code_complete_rnn, code_complete_LSTM} adopt statistical or corpus-based techniques like N-Gram, Deep Neural Network~\cite{lecun2015deep} (DNN), Recurrent Neural Network~\cite{rumelhart1986learning} (RNN), and Long Short-Term Memory~\cite{hochreiter1997long} (LSTM) models to learn program semantics. However, they do not easily generalize across domains and require expensive data collection, annotation, and training efforts to adapt to new tasks, limiting their real-world applicability. 

Recently, by leveraging pre-training, large language models (LLMs) have been largely adopted for code and shown capable of varied tasks including code completion ~\cite{codet5+, starcoder, roziere2023codellama}. 
These models, trained on large code corpora with trillions of tokens, can encode code knowledge and programming patterns into their large-scale parameters, remarkably outperforming the existing non-pre-trained techniques in both code understanding and generation tasks~\cite{zan2023large}.
For instance, \codellama{} ~\cite{roziere2023codellama} is capable of completing 53.7\% of the text-to-code programming problems upon the HumanEval~\cite{chen2021evaluating} benchmark with no additional adaptation. However, when faced with complex code logic or required to extrapolate beyond their training data, LLMs can struggle with incoherent or repetitive generations ~\cite{testcasedriven2021}. They may also hallucinate plausible but incorrect outputs ~\cite{llm_prob1, llm_prob2}. 
One potential solution to these issues is fine-tuning~\cite{alpaca, zheng2023judging, Xu2023WizardLMEL}, i.e., adapting a pre-trained language model to the code completion task by incrementally updating its parameters on the task-oriented dataset. 
However, this proves to be a costly endeavor both in terms of computational resources and time, making it impractical for many applications. For instance, fine-tuning the smallest LLaMA model typically requires a GPU workstation with eight A100 ~\cite{finetune_cost1, finetune_cost2}. 
On the other hand, the Retrieval Augmented Generation~\cite{meta2020rag} (RAG) techniques offer a more feasible solution. Specifically, in an RAG model, given an input, related information is retrieved from a pre-defined knowledge source and then used to assist the generation process. This retrieval process enables models to generate outputs that are coherent and relevant to the given context without the need for expensive fine-tuning. 
Accordingly, multiple code completion techniques~\cite{lu2022reacc, zhang2023repocoder, knm} have been proposed to leverage the power of RAG and have shown promising performance. 

Despite the promising results shown in the RAG-based techniques, they are still somewhat limited. First, they depend on extrinsic encoding models. Existing techniques construct code representations for retrieval mainly by encoding the code with auxiliary models, which requires additional efforts for training the encoding model. Second, the representational scope of these techniques is subject to a singular perspective on lexical semantics defined by the corresponding encoding model, i.e., they fail to account for the complexity and diversity inherent in code semantics.
To illustrate, for a complex missing line with rich context, 
only offering lexical semantics by the incomplete code is unlikely to fully represent the code intention. Under such a circumstance, we should retrieve code snippets that are most likely to produce analogous content, i.e., adopting more encoding perspectives other than lexical semantics only.
However, existing systems based on pre-defined encoding models require substantially additional training to encode the code from more perspectives, making it impractical for the existing RAG-based techniques. 
Therefore, it is essential to adopt a flexible retrieval approach to code context with multiple perspectives for code completion.





In this paper, we propose \ourmethod{}, a code completion framework leveraging prompt engineering and the contextual multi-armed bandit algorithm~\cite{li2010contextual} for the first time to flexibly incorporate and adapt to multiple perspectives of code. \ourmethod{} consists of two components---the \PromptRetrieval{} and the \PromptBandit{}. 
In particular, the \PromptRetrieval{} provides diverse perspectives of code while enabling flexible implementation and seamless integration with existing retrieval systems. Instead of creating a series of new embedding models with significant extra cost, we adopt prompt engineering which advances LLMs to understand code semantics via following human preferences and instructions~\cite{jiang-etal-2022-promptbert,jagerman2023query, Wang2023Query2docQE} 
such that we can access code semantics from different lenses for more comprehensive retrieval in a cost-effective manner. 
More specifically, our \PromptRetrieval{} examines three perspectives that are prominently used in RAG-based techniques, namely \CodeContext{}~\cite{lu2022reacc}, hypothetical line~\cite{zhang2023repocoder}, and code summarization~\cite{summary_generation}. We craft the prompt ``Embedding the following code snippets: [code]'' to encode the \CodeContext{}, and  ``<PRE> [Prefix] <SUF> [Suffix] <MID>'' to generate the hypothetical line serving as its representation, and ``This code snippets of [code] means'' to obtain the code summarization. To illustrate, ``[code]'' refers to the unfinished code, ``[Prefix]'' and ``[Suffix]'' represent the code snippets before and after the target insertion point. 
As our \PromptRetrieval{} presents three distinct perspectives, directly concatenating all three retrievals with input may overwhelm the completion model with misaligned perspectives.
To optimally select retrieved information with respect to the complex nature of incomplete code, we adopt the \PromptBandit{} based on a contextual multi-armed bandit algorithm where the different retrieval perspectives are seen as the ``arms'' of the bandit and the goal is to identify which arm (i.e., perspective) yields the highest reward or performance for individual incomplete code, conditioned on the similarity between retrieved snippets and incomplete code.
Accordingly, \ourmethod{} adapts to the dynamic nature of code completion, handles the uncertainty, and reliably determines the most suitable perspective for the LLMs to utilize in the code completion process.

We evaluate the effectiveness of \ourmethod{} on state-of-the-art (SOTA) models \codellama{}~\cite{roziere2023codellama} and \starcoder{}~\cite{starcoder}, extensively on 20 open-source repositories and 58 private-domain repositories from an e-commerce company with one billion Monthly Active Users. We also evaluate each component of \ourmethod{} and investigate how \ourmethod{} impacts the performance of fine-tuned models. Our evaluation results indicate that \ourmethod{} outperforms state-of-the-art code completion technique by 8.6\% on the open-source benchmark suite and 10.1\% on the private-domain benchmark suite in terms of \ExactMatch{}. Moreover, our prompt-based retrievers are robust across templates and comparable to external encoders. Furthermore, designing multiple retrieval perspectives (i.e., prompts) to elicit distinct semantic interpretations from the LLM allows us to obtain a wider range of code semantics. Incorporating the varied perspectives enriches the multifaceted representations, improving the code completion. \ourmethod{} also allows augmenting the fine-tuned techniques in a plug-and-play manner, yielding 5.6\% improvement over our studied fine-tuned model.



In summary, the contributions of this paper are listed as follows:
\begin{itemize}[leftmargin=*, topsep=0pt]
    \item \textbf{Novelty.} This paper opens up a new direction for multi-perspective code representation in retrieval-augmented code completion. We are the first to show incorporating prompt engineering and contextual multi-armed bandit can adapt to the most suitable code perspective without needing extra encoders. This provides a more comprehensive and adjustable encoding strategy compared to rigid representations in prior work~\cite{Shi2022CoCoSoDaEC, Jain2020ContrastiveCR, lu2022reacc}. 
    \item \textbf{Technique.} We propose and implement \ourmethod{} with two components, the \PromptRetrieval{} and the \PromptBandit{}. The \PromptRetrieval{} examines three essential perspectives via crafted instructions for code semantics.  
    The \PromptBandit{} dynamically chooses the most relevant perspective based on incomplete code context to provide the optimal contextual support for code completion. 
    \item \textbf{Evaluation.} We compare our approach to state-of-the-art techniques. \ourmethod{} outperforms the SOTA technique by 8.6\% on the open-source benchmark and 10.1\% on the private-domain benchmark in terms of \ExactMatch{}. We also show that our prompt-based retrievers are robust across templates and comparable to external encoders. Furthermore, designing prompts to elicit distinct semantic interpretations from the LLM allows us to obtain a wider range of code semantics. Incorporating the varied perspectives enriches the multifaceted representations, improving code completion. Finally, applying \ourmethod{} to fine-tuned models further produces a 5.6\% gain. 
\end{itemize}

%% file: sections/2_background.tex
\section{Background and Motivation}
\subsection{Code Completion}


Recent natural language processing (NLP) breakthroughs have facilitated large pre-trained language models for the code completion task~\cite{codet5+, nijkamp2023codegen}. In general, LLMs are built on the Transformer architecture and pre-trained on large-scale text corpora using self-attention mechanisms~\cite{Vaswani2017AttentionIA}. LLMs efficiently model contextual relationships and facilitate the learning of general linguistic interpretations. 
In particular, LLMs exhibit substantial model size and volume of training data. For instance, 
the smallest version of the LLaMA model~\cite{touvron2023llama}, launched in 2023, enables 7 billion parameters and is trained on 2 trillion tokens. 
LLMs predominantly adopt a decoder-only architecture, in which they aim to auto-regressively generate tokens based on all previously generated ones.

The training loss for typical LLMs, depicted in Equation \ref{eq1}, minimizes the negative log probability for the ground truth token $x_i$:
\begin{equation}
    \mathcal L = -\sum_i \log P_i (x_i|x_1,x_2,...,x_{i-1}; \theta)
    \label{eq1}
\end{equation}
where the conditional probability $P$ is modeled using a pre-trained language model $\mathcal{M}$ with parameters $\theta$. These parameters are optimized by applying the gradient descent algorithms~\cite{ruder2016overview} with respect to the input sequence $x_1,x_2,...,x_{i-1}$ preceding the given token $x_i$.

In particular, emerging code LLMs such as SantaCoder~\cite{Allal2023SantaCoderDR}, InCoder~\cite{InCoder}, \starcoder{}~\cite{starcoder}, and \codellama{}~\cite{roziere2023codellama} are trained using the Fill-in-the-Middle (FIM) objective~\cite{openaifim}. This technique involves randomly rearranging parts of a training sequence by moving them to the end and then generating predictions auto-regressively based on the reordered sequence. The pre-training loss for FIM remains consistent with Equation \ref{eq1}. Specifically for the code completion task, during inference, FIM leverages additional surrounding context by taking a prefix and suffix around the insertion point and generating the missing middle code tokens.
Formally, the goal is to generate the token $x_i$ that minimizes the negative log likelihood based on prefix tokens $[Prefix] = x_1,...,x_{i-1}$ and suffix tokens $[Suffix] = x_{i+1},...,x_j$ before and after the insertion point:
\begin{equation}
    \mathcal L = -\log P(x_i | [Prefix], [Suffix]; \theta)
    \label{eq2}
\end{equation}
For abbreviation, $\hat{X} = [Prefix], [Suffix]$ is denoted as the full unfinished code. Unlike conventional text generation models that are only conditional on preceding tokens, as formulated in Equation \ref{eq1}, access to suffix code is critical and practical for code completion. 
As a result, these infilling models largely outperform previous models in code completion. For instance, \codellama{} achieves an impressive pass@1 rate of 85.6\% on infilling benchmarks~\cite{InCoder}, surpassing previous SOTA techniques by a significant margin.

While LLMs have shown impressive capabilities on code completion tasks, they still face limitations when dealing with complex logic or are required to generalize beyond their training data~\cite{testcasedriven2021}. They may produce incoherent text when the generation requires long-term reasoning or even generate hallucinated outputs that seem plausible but do not actually reflect valid behaviors~\cite{llm_prob1, llm_prob2}. 



\begin{figure*}[h]
  \centering
  \includegraphics[width=0.85\linewidth]{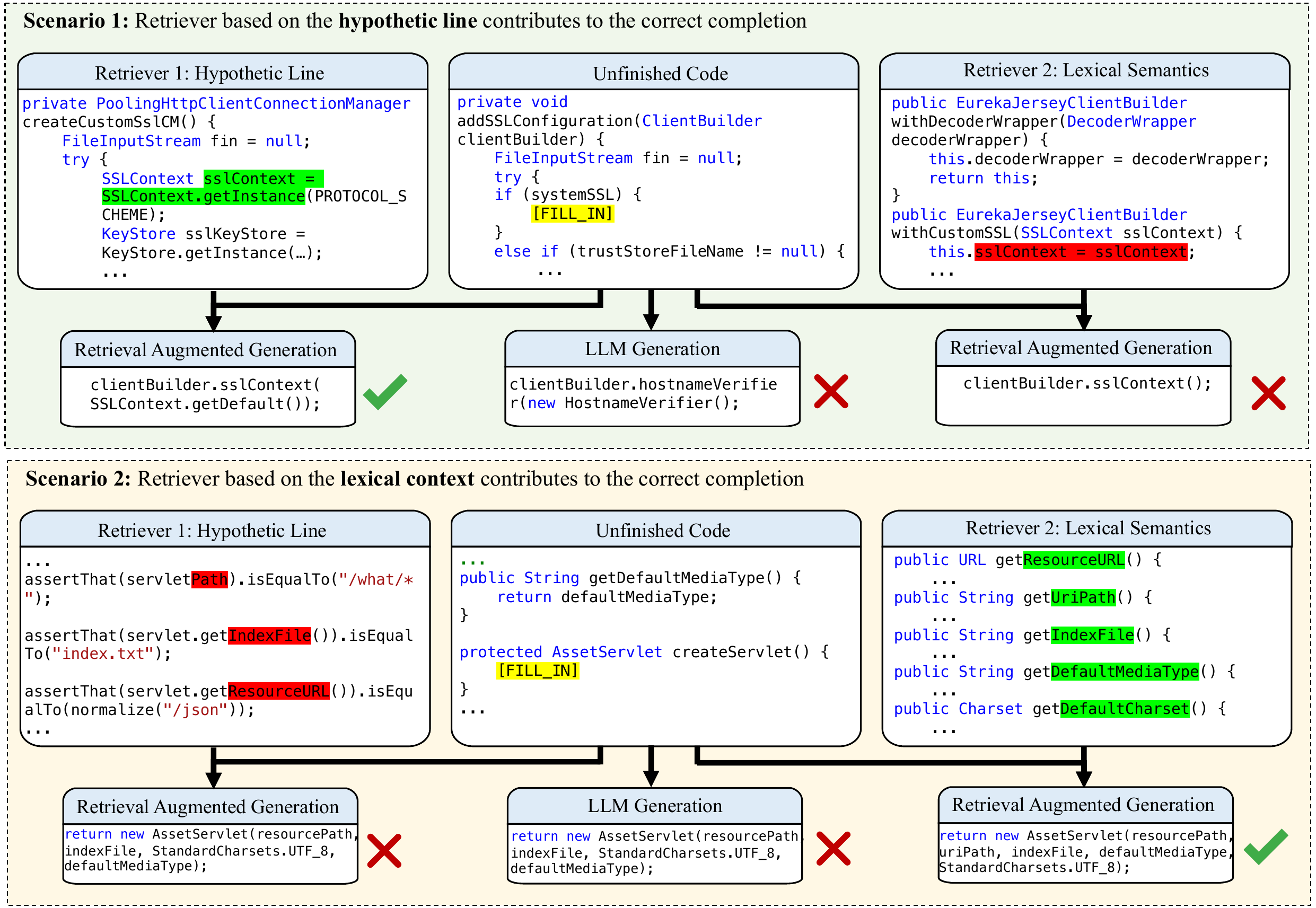}
  \caption{Code completion scenarios demonstrating the contextual dependence for optimal retrievals. 
  \colorbox{red}{\textcolor{black}{Red}} indicates the misleading information, \colorbox{green}{\textcolor{black}{Green}} represents the helpful hint.}
  \label{fig:intro-case}
\end{figure*}

\subsection{Retrieval-Augmented Generation}\label{rag background}
To address challenges on hallucination and enhance the production of coherent code, researchers have proposed that the models should be capable of accessing external memory or knowledge through information retrieval techniques, a.k.a. retrieval augmented generation (RAG)~\cite{meta2020rag}. 
The RAG process can be formulated as follows. First, the code database is split into snippets $C={c_1,c_2,...}$, which are encoded by the retrieval model $\mathcal{R}$ to derive their representations $h_{c_1}^{R},h_{c_2}^{R},...$ and form the corresponding database $D$. Second, for an incomplete snippet $\hat{X}$, the retrieval model $\mathcal{R}$ encodes it as $h_{\hat{X}}^{R}$ to retrieve relevant snippets $C^{R}$ based on distance functions between $h_{\hat{X}}^{R}$ and the representations $h_{c_1}^{R},h_{c_2}^{R},...$ stored in database $D$. 
The retrieval process can be described as:
\begin{equation}
    p_{R}(C|x_1,x_2,...,x_{i-1},D)
    \label{eq3}
\end{equation}
The retrieved context, denoted as $C^{R}$, along with unfinished code $\hat{X}$ are then consumed by the model to conduct the code completion:
\begin{equation}
    p_{\theta} (x_i|\hat{X}, C^{R})
    \label{eq4}
\end{equation}
Consequently, the retrieved code snippets can be perceived as supplemental knowledge and interpreted by the LLMs to facilitate coherent generation. 

There are two primary strategies for RAG---per-token and per-output~\cite{meta2020rag}. In per-token RAG (RAG-token), distinct code snippets are assigned to individual tokens, whereby each new token generation requires the retrieval of a new code snippet. Therefore, this strategy escalates the retrieval and the encoding demands proportionally with the length of the generated sequence, leading to increased computational time. Moreover, the necessity for accessing each stage of token generation constrains its integration with closed-source systems, such as GPT~\cite{openai2023gpt4}, which operate as black boxes, i.e., exposing no intermediate generations.
In contrast, per-output RAG (RAG-sequence) leverages the same code snippet to conduct the generation of the entire sequence, necessitating only one retrieval phase prior to generation. This strategy not only enhances efficiency of the retrieval process but also facilitates a more streamlined integration into the existing frameworks. Consequently, our subsequent discussions and analyses will be exclusively concentrated on the per-output RAG.

Following the RAG-sequence framework, the pioneering work ReACC~\cite{lu2022reacc} utilizes a dual-encoder model to function as a code-to-code search retriever and employs an auto-regressive language model to execute code completion. RepoCoder~\cite{zhang2023repocoder} suggests refining the retrieval process by iteratively utilizing the most recently generated content to retrieve information. 


\subsection{Motivation} 
As discussed in Section~\ref{sec:intro}, the existing RAG-based techniques rely on external encoding models and are constrained to a singular perspective defined by the corresponding encoding model. They fail to account for the complexity and diversity inherent in code semantics.

Figure~\ref{fig:intro-case} presents two distinct code completion scenarios that illustarte the importance of different retrieval perspectives. In Scenario~1, Retriever~1 provides the relationship between ``sslContext'' and ``SSLContext.getInstance()'' which gives relevant context to assist the generator in completing the line. While Retriever~2 also hints ``sslContext()'', it is insufficient to bridge the gap for completing the ``SSLContext.getDefault()''. Conversely, in Scenario~2, Retriever~2 presents the list of parameters, i.e., ``ResourceURL, UriPath, IndexFile, DefaultMediaType, DefaultCharset'', that should be included in the function ``createServlet()''. This can be easily integrated by the generator to complete the return line ``return new AssetServlet(resourcePath, uriPath, indexFile, defaultMediaType, StandardCharsets.UTF\_8);'', whereas Retriever~1 presents noisy context and hinders the correct code completion. These scenarios indicate the need to retrieve the code from multiple perspectives and select optimal retrieval. Thus, the complex nature of incomplete code poses the need for adjustable encoding perspectives to cover as much code semantics as possible. However, existing systems rely on pre-defined encoding models and require substantially additional training to encode the code from more perspectives. Thus they are limited in adjusting the retrieval perspective to cope with multifaceted code semantics. 
Accordingly, we can infer that it is essential for a flexible retrieval approach to adapt to code context with multiple perspectives for code completion.

In this paper, we attempt to address the limitations of prior code completion techniques by taking a more flexible approach based on \PromptRetrieval{} and \PromptBandit{} (as illustrated later). First, we leverage prompt engineering techniques to elicit a deeper, multi-faceted interpretation of the incomplete code based on LLM. By crafting prompts that guide the LLM to focus on multiple perspectives, 
we can expand the code representation beyond lexical features with no extra cost to train additional encoding models. This allows us to retrieve code snippets that can hint the completion even upon lexical dissimilarity. Second, we could attempt to employ adaptive selection to choose the most suitable retrieved results, i.e., dynamically making decisions from different retrieved information based on the specifics of the semantics of the target incomplete code snippet. 


%% file: sections/3_approach.tex
\begin{figure*}[ht]
  \centering
  \includegraphics[width=0.80\linewidth]{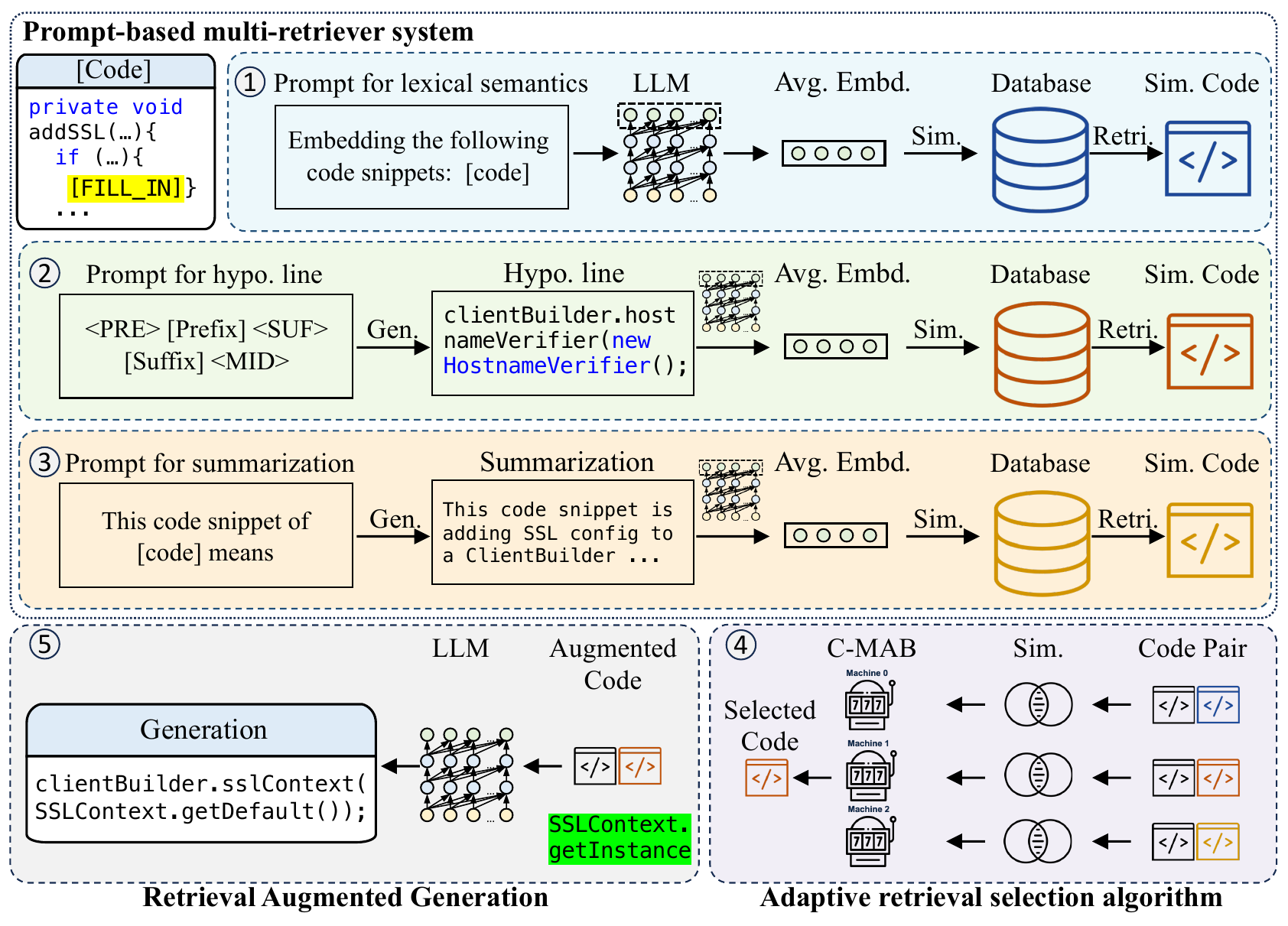}
  \caption{The \ourmethod{} framework. The \PromptRetrieval{} encodes the \CodeContext{} \ding{172}, hypothetical line \ding{173}, and code summarization \ding{174} to derive multi-perspective representations. The \PromptBandit{} \ding{175} makes decisions based on code semantics similarities and selects the optimal context from retrievals. Finally, the selected code is concatenated with the unfinished code for augmented generation \ding{176}.}
  \label{fig:prompt_embedding}
\end{figure*}

\section{Approach}
\subsection{Overview}
In this paper, we introduce \ourmethod{}, a novel framework leveraging prompt engineering and the contextual multi-armed bandit algorithm~\cite{li2010contextual} to select suitable perspectives for code completion. As shown in Figure \ref{fig:prompt_embedding}, \ourmethod{} adheres to the Retrieval-Augmented Generation (RAG) framework, as outlined in Equation \ref{eq3} (the retrieval phase) and Equation \ref{eq4} (the augmented generation phase).
\ourmethod{} consists of two components---the \PromptRetrieval{} and the \PromptBandit{}. 
In particular, by adopting prompt engineering, the \PromptRetrieval{} encodes the \CodeContext{}, hypothetical line, and code summarization to derive multi-perspective representations. More specifically, we employ three prompt retrieval models $\mathcal{R}_1$, $\mathcal{R}_2$, $\mathcal{R}_3$ that encode the \CodeContext{} (\ding{172}), generate a hypothetical line (\ding{173}), and produce the code summarization (\ding{174}) respectively. This allows retrieving relevant snippets $C^{R_1}, C^{R_2}, C^{R_3}$ from the database based on representation similarities.
Moreover, we use the contextual multi-armed bandit algorithm (\ding{175}) to make decisions based on code semantics and select the optimal context from retrieved $C^{R_1}, C^{R_2}, C^{R_3}$. Note that we are the first to introduce such an adaptive retrieval selection mechanism for code completion.
The selected retrieval, for example, $C^{R_1}$, along with unfinished code $\hat{X}$ is then consumed by the model to perform the code completion, i.e. $P(x_1|\hat{X}, C_{R_1};\theta)$ (\ding{176}). 


\subsection{Prompt-based Multi-Retriever System} 
We formulate the multi-retriever in a unified paradigm by prompting the LLMs with designed prompt templates (or prompts for short for the rest of the paper). 
The construction of the retriever in the existing techniques is performed by encoding the code using auxiliary models, which requires additional resources for training the encoding model.
By leveraging the LLM knowledge and crafting prompts, we can seamlessly represent code semantics from diverse perspectives with no need for extra models. This unified paradigm offers the advantage of flexibility and efficiency, thereby simplifying the process of constructing a multi-retriever system. Formally, given a language model $\mathcal{M}$, incomplete snippet $\hat{X}$, and a crafted prompt $Prompt$, we execute the model to process the input as Equation~\ref{eq5}:
\begin{equation}
    Out = \mathcal{M}(Prompt;\hat{X})
    \label{eq5}
\end{equation}
We then extract the corresponding hidden states $h$ of the output $Out$ as the representation for the target perspective of the code $\hat{X}$. Finally, by crafting the prompts towards the \CodeContext{}, hypothetical line, and code summarization perspectives, we construct the following three retrievers.

\paragraph{Lexical Semantics.}\label{sec:code_context}
Incorporating lexical semantics into the retrieval process allows us to fetch relevant code snippets in terms of lexical semantics. Conventional techniques leverage contrastive pre-training to learn the code similarity~\cite{Shi2022CoCoSoDaEC, Jain2020ContrastiveCR, lu2022reacc}. For example, ContraCode~\cite{Jain2020ContrastiveCR} employs pre-training of an LLM to differentiate functionally similar program variants against non-equivalent distractors. However, in the domain of code completion, the code is typically unfinished and may not express coherent or consistent meaning, which is significantly different from the training samples of these contrastive models. Note that for code LLMs, as in Equation~\ref{eq2}, they are pre-trained to auto-regressively generate the next token and complete the code. Hence, we deeply explore the knowledge embedded within LLMs by crafting prompts to encode incomplete snippets, as in Figure \ref{fig:prompt_embedding} \ding{172}. In particular, we craft the prompt, ``Embedding the following code snippets: [code]'' to encode the \CodeContext{}, where the ``[code]'' refers to the unfinished code. Then we extract the last hidden layer for the whole prompt and average them as the representation for \CodeContext{}.

\paragraph{Hypothetical Line.}

Hypothetical line refers to the potential line that can complete the code, motivated by the concept of Hypothetical Document Embeddings (HyDE)~\cite{hypothetical}. In HyDE, a query question is passed into the model and guided to ``write a document that answers the question'', leading to the generation of a hypothetical document. This hypothetical document is then transformed into an embedding in a vector space, enabling database search for retrieval. 
In our \PromptRetrieval{}, as illustrated in Figure \ref{fig:prompt_embedding} \ding{173}, we use the prompt ``<PRE> [Prefix] <SUF> [Suffix] <MID>'' to generate the hypothetical line that acts as its representation where ``[Prefix]'' and ``[Suffix]'' represent the code snippets before and after the target insertion point respectively. Note that this structure is consistent with the pre-training format for \codellama{}. For the code snippet $C$, we mask the line and input the surrounding code into the model $\mathcal{M}$ to generate the hypothetical line, storing the corresponding embedding to build the retrieval database $D$. For each incomplete code snippet $\hat{X}$, we follow the same process to generate its embedding for database search. 
Employing hypothetical line representations provides a critical benefit for retrieval---enriching the representation of incomplete code beyond the surrounding context. Generating embeddings solely from the available \CodeContext{} is prone to lack important semantics and patterns contained in the missing line itself. On the other hand, by prompting the model to generate a hypothetical line, the produced text exhibits relevant attributes like variable names, data types, and function signatures even when the \CodeContext{} alone does not offer such information. In other words, if two incomplete snippets generate hypothetical completions with similar or analogous variable names, function calls, etc., their overall functionality can be likely similar. 
This facilitates retrieving code snippet with  conceptually relevant but lexically dissimilar information to the incomplete code.


\paragraph{Code Summarization.}

Code summarization allows capturing the overall functionality and the purpose of a code snippet in natural language. As humans tend to reason about code at a higher level of abstraction, summarization embeddings allow retrieval to focus on semantic similarity rather than superficial syntactic matches.
Furthermore, natural language summaries provide a mechanism to inject human preferences into representation learning. 
This allows retrievals to better match human judgments of conceptual similarity.
In our \PromptRetrieval{}, as shown in Figure~\ref{fig:prompt_embedding} \ding{174}, we use ``This code snippet of [code] means'' to produce code summarization and average the summary embeddings as the representation for the code.

In fact, our prompt engineering can be easily extended beyond these perspectives, i.e., it can readily expand to encode any semantic dimensions of interest, which makes our \PromptRetrieval{} potentially robust in the real world. 

\subsection{Adaptive Retrieval Selection Algorithm}
After gathering retrieval information from multiple retrievers, it is essential to determine the optimal hints that aid code completion. 
Specifically, given the varied coverage and perspectives of different retrievers, directly concatenating all of them with the input could cause information overload and perspective confusion for code completion.
Therefore, we aim to dynamically tailor the selection of the most suitable perspective for incomplete code. 

We propose to tackle this perspective selection challenge as a learning problem~\cite{bandits_multilingual}. The learning agent is presented with an action space that includes \CodeContext{}, hypothetical line, and code summarization perspectives. The agent's objective is to identify and select the most fitting prompt perspective while receiving rewards that reflect the quality of the selected perspective. To implement this learning approach, we adopt the multi-armed bandit algorithm, which is widely employed for recommendation systems~\cite{bandit_rec1, bandit_rec2}. Specifically, we use the LinUCB algorithm~\cite{chu2011contextual}, a variant of the multi-armed bandit algorithm that takes context into account. 
The LinUCB algorithm offers the following advantages: 1) it enhances decision-making by continuously learning from outcomes and refining the selection process, and 2) it adapts to different contexts, optimizing performance for each input. 
As illustrated in Figure~\ref{fig:prompt_embedding} \ding{175}, the different retrieval perspectives are seen as the "arms" of the LinUCB and the goal is to identify which arm (i.e., perspective) yields the highest reward for each individual incomplete code snippet conditioned on the context.
We choose the similarity score from each retriever as one dimension the context for our LinUCB algorithm. Additionally, we use the Jaccard similarity between the retrieved snippets and incomplete code as another dimension of context for our LinUCB algorithm, following the Jaccard metrics~\cite{jaccard1912distribution} used in RepoCoder~\cite{zhang2023repocoder}. 
We also set the \ExactMatch{} score as the reward for the LinUCB algorithm where the reward is 1 upon an \ExactMatch{} and 0 otherwise.
Note that the algorithm is flexible and can be readily extended to include additional dimensions, such as other possible retrieval perspectives, techniques or reward metrics.
As a result, 
the LinUCB algorithm makes informed decisions about the optimal perspective for each incomplete code snippet.

\begin{equation}
\begin{aligned}
    A &\leftarrow A + x_{i,a}x_{i,a}^\top \\
    b &\leftarrow b + x_{i,a}r_i \\
    \text{retrieval} &= \arg\max_a \left( A^{-1}bx_{a} \right)
\end{aligned}
\label{select_arm_eq}
\end{equation}

As described in Equation~\ref{select_arm_eq}, we first train the parameters \( A \) and \( b \) of the LinUCB algorithm. Specifically, the matrix \( A \), which accumulates the outer products of feature vectors for the samples $i$, is updated as \( A + x_{i,a}x_{i,a}^\top \), while the vector \( b \), aggregating the product of rewards and feature vectors, is updated as \( b + x_{i,a}r_i \). Here, \( x \) denotes the feature vector of the current arm, incorporating the Jaccard and cosine similarities between the incomplete code and the retrieval results, $r_i$ refers to the reward reflecting the \ExactMatch{} result in our situation.
Finally, the selected retrieval from the LinUCB algorithm along with the unfinished code is then consumed by the model to conduct the code completion (\ding{176}). 

%% file: sections/4_evaluation.tex
\section{Evaluation}

To evaluate \ourmethod{}, we have formulated the following three research questions:

\begin{itemize}[leftmargin=*]
    \item \textbf{RQ1}: How does \ourmethod{} perform compared with state-of-the-art code completion techniques? 
    \item \textbf{RQ2}: How do individual components of \ourmethod{} impact the performance? 
    \item \textbf{RQ3}: How does \ourmethod{} perform compared with fine-tuning? Can it further improve a fine-tuned model? 
\end{itemize}

\noindent

\subsection{Experiment Setup}\label{sec:setup}
\subsubsection{Models}
We choose two latest state-of-the-art code LLMs as the base models in our paper.
\begin{itemize}[leftmargin=*]
    \item \textbf{\starcoder{}~\cite{starcoder}}, released on May 2023, is trained on 1 trillion tokens from GitHub, including 80+ programming languages, Git commits, GitHub issues, and Jupyter notebooks. In particular, we use StarCoderBase 15.5B for the code completion task.
    \item \textbf{\codellama{}~\cite{roziere2023codellama}}, released on August 2023, is continuously trained on extensive 500B code data based on the Llama 2 model~\cite{touvron2023llama}. It possesses state-of-the-art capabilities in code comprehension. Specifically, we use \codellama{} 13b-Instruct for embedding and generation. 
\end{itemize}

\input{tables/main_results}

\subsubsection{Baselines}
We adopt the baseline techniques as follows. Specifically, we choose the code completion frameworks based on the RAG-sequence strategy only with as stated in Section \ref{rag background}. 

\begin{itemize}[leftmargin=*]
    \item \textbf{BM25~\cite{parvez2021retrieval}.} BM25 is proposed upon the BM25 ranking algorithm~\cite{robertson2009probabilistic}, which is one of the most widely employed retrieval algorithms in the Question and Answering (QA) domain. We leverage BM25 to search for code similar to the given incomplete code. The retrieved code is then concatenated with the incomplete code and input into the LLMs for completion.
    \item \textbf{ReACC~\cite{lu2022reacc}.} ReACC, published in ACL2022, employs the vanilla RAG framework for code completion. Since it does not provide complete reproducible encoding models for retrieval, we implement the framework by using the latest code retrieval model GTE-large~\cite{gte_model}, which was released on August 2023 and achieves the SOTA performance for now on code search.
    \item \textbf{RepoCoder~\cite{zhang2023repocoder}.} RepoCoder, published in ENMLP2023, refines the code retrieval process by iteratively utilizing the most recently generated content to retrieve information. Following the original setting, we reproduce the algorithm using UniXcoder~\cite{Guo2022UniXcoderUC} as its embedding model.
\end{itemize}

\input{tables/testsets}

\subsubsection{Benchmark Construction}\label{sec:benchmark}
To comprehensively evaluate the performance of LLMs in code completion scenarios, we follow the protocol of previous work~\cite{zhang2023repocoder} to crawl 20 high-quality code repositories from GitHub, covering multiple levels of code completion---random line completion and function body completion. These scenarios, often encountered by developers, can largely reflect real-world development scenarios. We randomly split 10\% of the files as the test set, 10\% for validation used in training the \PromptBandit{}, 
and the rest as retrieval data. Following ~\cite{zhang2023repocoder}, for the line completion, we randomly select three lines from each test file as test cases. 
For the function body completion, we extract all functions in test files.
Eventually, we obtain a test dataset with 3317 instances. We conduct the same process for the validation set. 

Given that LLMs are pre-trained on expanded GitHub datasets, it might inadvertently encompass elements from our test set and lead to the risk of test set contamination. To alleviate this issue, we construct another benchmark based on private-domain code from an e-commerce company with around one billion Monthly Active Users. We collect 58 repositories and construct the dataset with the same protocol as discussed above. In total, we construct a test dataset with 3074 instances. Table~\ref{tab:test_datasets} shows the test set details.


\subsubsection{Metrics}
Following previous studies~\cite{zhang2023repocoder, lu2022reacc}, we select two widely recognized evaluation metrics for code generation---\textit{Exact Match (EM)} and \textit{Edit Similarity (ES)}. In particular, EM quantifies the percentage of generated code snippets that exactly match the ground truth. 
ES, adapted from the Levenshtein Edit Distance~\cite{levenshtein1965binary}, measures the required edit operations from generated content to the ground truth. 

\subsubsection{Implementation}
We use the Python implementation of the \starcoder{}-15.5B and the \codellama{}-13b-Instruct models obtained on Hugging Face~\cite{wolf2019huggingface}. Note that the models serve a dual purpose---they are employed for encoding the incomplete code from various perspectives through prompt engineering, and concurrently, for executing the code completion task.
We implement the Dense vector retrieval using \textit{Faiss}~\cite{faiss}. For the LinUCB algorithm, we set the coefficient of the upper confidence bound $\alpha = 0.1$. We conduct LLM generation using the \textit{Text-Generation-Inference} framework~\cite{tgi} with greedy decoding to minimize randomness. For fine-tuning (Section~\ref{sec:finetune}), we set $batch\:size=12$ and $learning\;rate=2e{-}5$ and train the models with the AdamW optimizer~\cite{loshchilov2019decoupled} for 2 epochs\footnote{We evaluate the impact of multiple reasonable training setups and present the results in our GitHub repository~\cite{procc} due to page limit. Note that applying these setups does not incur significant performance variations.}. 
To ensure fairness in the analysis of time and space complexity, all experiments are performed on a cluster equipped with 8 NVIDIA A100-80GB GPUs.  

%% file: tables/main_results.tex
\begin{table*}[!ht]
        \centering
        \caption{Main results on open-source benchmark. Numbers are shown in percentage (\%)}
        \label{tab:main_results}
        \small 
        \setlength\tabcolsep{4pt} 
        \begin{adjustbox}{width=1.9\columnwidth}
        \begin{tabular}{cp{1.2cm}p{1.2cm}p{1.2cm}p{1.2cm}p{1.2cm}p{1.2cm}p{1.2cm}p{1.2cm}p{1.2cm}}
        \hline
        Metrics & Type & Base & BM25 & RepoCoder & ReAcc & Lexical Semantics & Code Summary & Hypo. Line & \ourmethod{} \\ \hline
        \multirow{3}{*}{\makecell{Code Llama \\ EM}} 
                & FB.  & 34.97 & 36.34 & 37.35 & 37.12 & 37.21 & 38.97 & 37.90 & \textbf{41.16} \\
                & RL.  & 68.91 & 71.44 & 71.37 & 71.20 & 71.91 & 71.60 & 73.89 & \textbf{76.58} \\
                & Avg. & 47.90 & 49.71 & 50.31 & 50.11 & 50.44 & 51.40 & 51.61 & \textbf{54.66} \\ 
         \hline
        \multirow{3}{*}{\makecell{Code Llama \\ ES}} 
                & FB.  & 64.49 & 64.85 & 65.77 & 65.53 & 65.21 & 66.33 & 64.85 & \textbf{67.22} \\
                & RL.  & 87.15 & 88.41 & 87.82 & 87.81 & 87.82 & 87.80 & 88.88 & \textbf{89.86} \\
                & Avg. & 73.12 & 73.83 & 74.17 & 74.02 & 73.83 & 74.55 & 74.01 & \textbf{75.85} \\ 
         \hline
        \multirow{3}{*}{\makecell{StarCoder \\ EM}} 
                & FB.  & 30.83 & 30.05 & 30.63 & 30.59 & 31.55 & 32.23 & 31.49 & \textbf{33.85} \\
                & RL.  & 68.04 & 72.15 & 71.26 & 71.20 & 71.47 & 71.72 & 72.23 & \textbf{73.97} \\
                & Avg. & 45.01 & 46.10 & 46.11 & 46.07 & 46.76 & 47.28 & 47.01 & \textbf{49.14} \\ 
        \hline
        \multirow{3}{*}{\makecell{StarCoder \\ ES}} 
                & FB.  & 60.56 & 59.39 & 59.78 & 59.67 & 59.90 & 61.23 & 61.10 & \textbf{62.93} \\
                & RL.  & 86.99 & 88.53 & 88.23 & 88.20 & 87.37 & 87.12 & 87.27 & \textbf{88.53} \\
                & Avg. & 70.63 & 70.50 & 70.62 & 70.54 & 70.37 & 71.10 & 71.07 & \textbf{72.69} \\ 
        \hline
        \end{tabular}
        \end{adjustbox}
\end{table*}
        

%% file: tables/testsets.tex
\begin{table}[t]
\centering
\caption{Statistics of test datasets}
\label{tab:test_datasets}
\begin{tabular}{llll}
\hline
\textbf{Domain} & \textbf{Type} & \textbf{Abb.} & \textbf{Count}  \\ 
\hline
\multirow{3}{*}{\makecell{Open-Source}}
& Function Body & FB. & 2053 \\ 
& Random Lines & RL. & 1264 \\ 
& ALL & - & 3317 \\
\hline
\multirow{3}{*}{\makecell{Private-Domain}}
& Function Body & FB. & 1972 \\ 
& Random Lines & RL. & 1102 \\ 
& ALL & - & 3074 \\
\hline
\end{tabular}
\end{table}

%% file: sections/5_results.tex
\subsection{Results and Analysis}  


\subsubsection{RQ1}
Table~\ref{tab:main_results} presents the performance comparison results between \ourmethod{} and the studied baseline techniques on top of the open-source benchmark suite, where ``FB'' refers to function body, ``RL'' refers to random line, and ``Avg'' refers to averaged results. We can observe that \ourmethod{} can significantly outperform the baseline techniques. In particular, for the average results on \codellama{}, \ourmethod{} significantly improves the code completion task with 14.1\% EM improvement over the vanilla model (from 47.90 to 54.66) and demonstrates an 8.6\% EM improvement over the previous SOTA technique RepoCoder (from 50.31 to 54.66). 
For the average results on \starcoder{}, \ourmethod{} achieves 9.2\% and 6.6\% EM gain over the vanilla model and RepoCoder respectively (from 45.01/46.11 to 49.14).
These results indicate the power of \ourmethod{} for effectively retrieving relevant contextual information to enhance code completion effectiveness. 
When examining the types of code completions, \ourmethod{} shows remarkable performance in both multiple-line and function body completion scenarios, having a 17.7\% EM improvement over the vanilla code completion technique (from 34.97 to 41.16). This suggests that \ourmethod{} is particularly effective at handling more complex code completion scenarios. 

Table~\ref{table:main_results_kwai} summarizes the experimental results for our private-domain benchmark suite. Within this benchmark suite, the baseline \codellama{} model achieves an average EM score of 33.67, in contrast to the 47.90 EM observed in the open-source benchmark. This difference indicates that LLMs are in general limited in dealing with the domain-specific task where the test data fall outside their training corpus. Interestingly in this scenario, \ourmethod{} has demonstrated an even more significant enhancement compared with our evaluations on the open-source benchmark suite, e.g., having a 36.8\% EM increase over the baseline model (from 33.67 to 46.07) and surpassing the prior SOTA BM25 by 10.1\% (from 41.84 to 46.07). 

\input{tables/main_results_kwai}

Our investigation reveals that each component of \ourmethod{}, i.e., the individual prompt-based retrievers for \CodeContext{}, hypothetical line, and code summarization, attain results comparable with prior SOTA approaches that rely on additional embedding models. 
For instance, for the average results with \codellama{} on the open-source benchmark suite, they achieve 50.44, 51.61, and 51.40 EM respectively, which all slightly outperform 50.31 for RepoCoder. While on the private-domain benchmark suite, they obtain 41.42, 42.79, and 42.10 EM respectively, which also all outperform 41.70 for RepoCoder.
These findings affirm the power of our prompt-based retrievers to effectively harness the knowledge embedded within LLMs. 


\subsubsection{RQ2}\label{sec:rq2}
In this section, we systematically evaluate the effects of the key components in \ourmethod{}. For simplification, all the experiments are conducted based on \codellama{}.

\begin{figure}[t]
  \centering
  \includegraphics[width=0.9\linewidth]{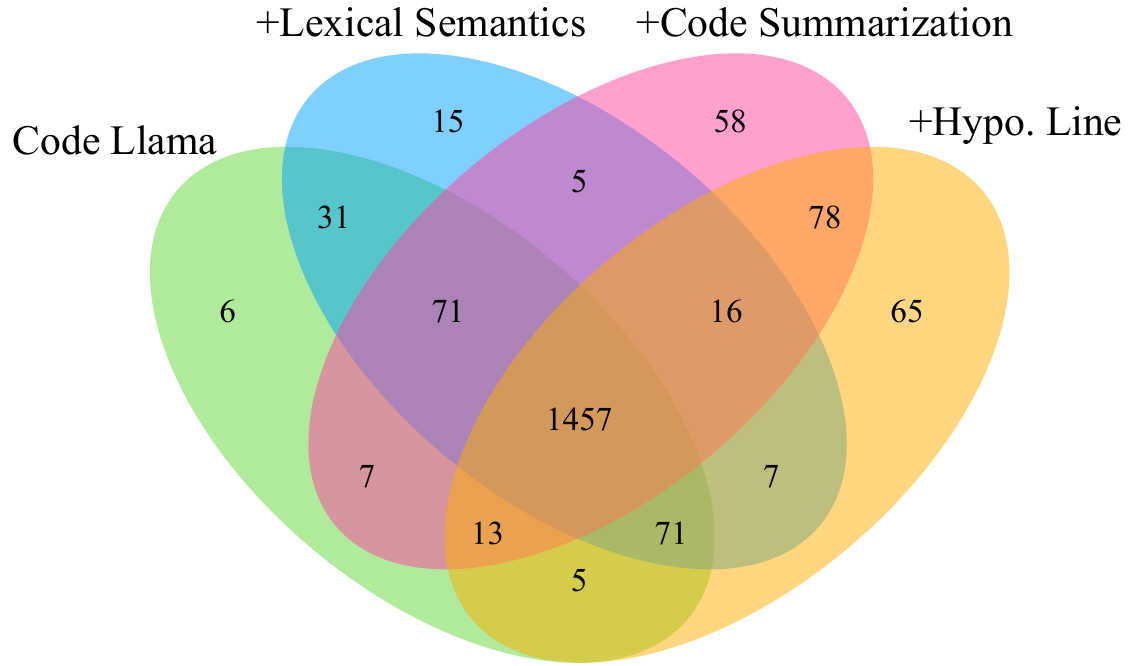}
  \caption{Venn diagram of different retrievers. It shows the number of samples that are completed correctly in the open-source benchmark.}
  \label{fig:prompt_venn}
\end{figure}

\paragraph{Prompting Perspectives}
Our default individual retrievers focus on three perspectives---\CodeContext{}, hypothetical line, and code summarization. 
Specifically, we plot a Venn diagram of the unique \ExactMatch{} achieved by each prompting template in Figure~\ref{fig:prompt_venn}. It is observed that each template brings to light different aspects of code semantics, thereby achieving exclusive successes in their targeted perspectives. For example, Table~\ref{table:prompting_template} shows that Prompt No.3 uniquely retrieves correct contextual hints to complete 65 samples, demonstrating specialized strengths in its particular facet. 

We further involved an evaluation of potential facets by designing various prompting perspectives to encode semantics from distinct perspectives. All the prompt perspectives utilized are presented in Table~\ref{table:prompting_template} where ``[code]'' symbolizes the incomplete code snippet, structured according to the FIM paradigm as described in Equation~\ref{eq2}. The placeholder ``[generation]'' signifies the output generated by the LLM conditioned on the corresponding prompt template. We derive the average embedding from the last hidden layer of LLM for prompt or generation, which forms the basis for the code representation in the retrieval process. Note that our default prompts are No.1, No.3, and No.5 in Table~\ref{table:prompting_template}. Our observations indicate that the directive prompts are effective and achieve comparative performance (between 50.41 to 51.61) with external encoding models (between 49.71 to 50.31) as in Table~\ref{tab:main_results}. Note that the construction of prompt perspectives only introduces minor variability in the code completion performance. For instance, a small variation like substituting ``Embedding'' with ``Representing'' in Prompt No.1 leads to negligible 0.03 absolute EM differences. 
Nevertheless, the overall robustness and effectiveness of code completion tasks are consistently maintained.

In conclusion, the capability to understand and generate informative representations solely from prompt instructions highlights LLMs' inherited alignment with human intents. By carefully formulating prompts for specific facets, we can readily exploit LLMs' knowledge to serve retrieval goals. The effectiveness indicates prompt engineering's potential as a lightweight yet powerful technique for multifaceted encoding.

\input{tables/prompting_template}


\paragraph{\PromptBandit{}}

Considering the prohibitive complexity of evaluating all combinations of prompting perspectives, our investigation on the \PromptBandit{} is confined to include the perspectives and combinations between the most distinctive perspectives, i.e., prompts No.1, No.3, and No.5 in Table~\ref{table:prompting_template}. Table~\ref{table:retrieval_selection_retriever} presents the evaluation results which reveal that combining various perspectives outperforms individual ones, showcasing the value of multi-perspective retrieval.
For instance, combinations of Prompt No.1+3, No.1+5, and No.3+5 yield EM scores of 52.43, 52.94, and 53.51 respectively, while the corresponding best single perspective retrieval presents an EM score of 51.61 (Prompt No.3).
Additionally, incorporating all six semantic perspectives from Table~\ref{table:prompting_template} further improves the code completion performance, though gains are marginal with only 0.22 absolute EM gain over 54.66 EM from the combination of three prompts No.1, 3, and 5. This slight improvement indicates that while expanding a wide range of perspectives can offer benefits, potential overlapping information can limit its improvement. In conclusion, our strategic prompting approach from three perspectives elicits distinct semantic interpretations from the LLM to obtain a wider range of code semantics. Incorporating these varied perspectives enriches the multifaceted representations and contributes to improved code completion performance.


\input{tables/retrieval_selection_retrievers}

Furthermore, we examine multiple selection algorithms for the multi-retrievers. we first directly concatenate all three retrievals with input as one compared technique, namely ``Union''. Then we apply a naive selection approach that selects the prompting retriever based on the maximum dense vector similarity, called ``Max Similarity''. We also re-frame this selection task as a classification problem, implementing logistic regression to choose the appropriate prompting retriever, called ``Logistic Regression''. At last, we include the optimal single perspective retriever with the hypothetical line as a reference.
Table~\ref{table:retrieval_selection} shows the detailed results.
Our findings indicate that directly concatenating all three retrievals with input achieves 51.97 EM. Direct concatenating risks presenting excessive information and only provides 0.36 slight absolute improvement over the best single perspective retrieval with the hypothetical line, which achieves 51.61 EM. Employing the maximum similarity and logistic regression techniques generally enhances the performance of the best single retriever with 0.76 and 1.6 absolute EM improvement. 
The LinUCB algorithm, leveraging the similarity scores as contextual information, achieves the best performance with a 3.05 absolute EM improvement over the best single retriever and makes decisions about the optimal perspective for incomplete code. 

\input{tables/retrieval_selection}

\subsubsection{RQ3}\label{sec:finetune}
This section presents a systematic comparison between \ourmethod{} and the conventional fine-tuning approach, evaluating their advantages and limitations. Moreover, we investigate the potential of \ourmethod{} to enhance the performance of models that have been already fine-tuned, thereby assessing its value as a supplementary optimization technique for fine-tuned models. For simplification, all the experiments are conducted based on \codellama{}.


\paragraph{Training Dataset}
For a fair comparison, the test dataset employed for fine-tuning is identical to that used for retrieval, i.e., a total of 3317 test samples from a corpus of 20 repositories. The remaining files within these repositories are utilized to construct the validation and training sets, following the same protocol outlined in Section~\ref{sec:benchmark}. This process yields 26,838 training samples used for the fine-tuning of hyper-parameters and 3156 validation samples.

\begin{figure}[b]
  \centering
  \includegraphics[width=0.95\linewidth]{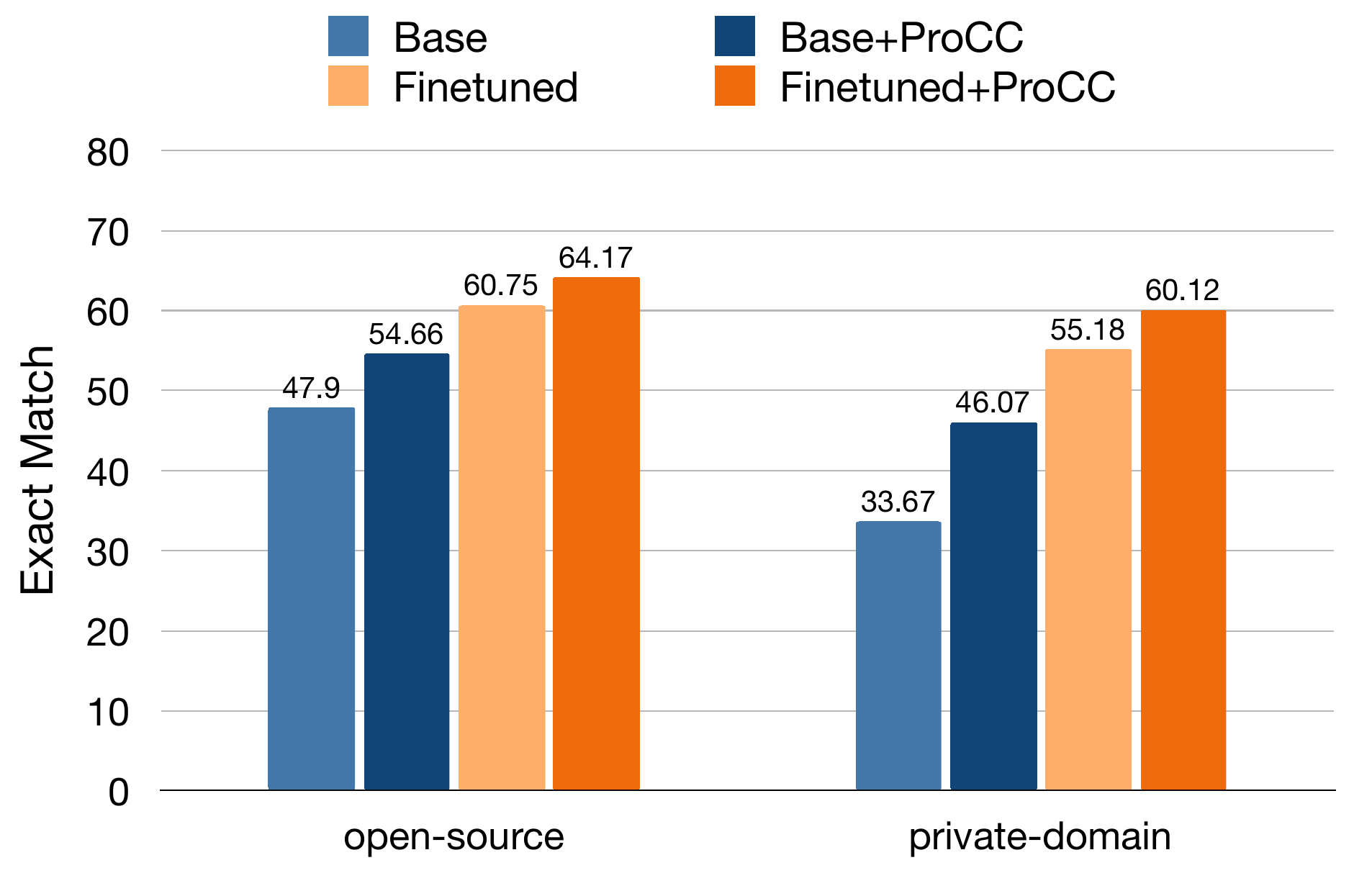}
  \caption{Finetune v.s. \ourmethod{}}
  \label{fig:rq3}
\end{figure}

\paragraph{Results}
In addition to the fine-tuning experiment, we apply \ourmethod{} to the fine-tuned models. Figure~\ref{fig:rq3} presents the evaluation results which show that fine-tuning significantly enhances code completion performance, achieving an absolute 12.85 EM gain on the open-source benchmark suite and 21.51 EM on the private-domain benchmark suite compared to the baseline model respectively (from 47.9 to 60.75 and from 33.67 to 55.18). Furthermore, the application of \ourmethod{} to this fine-tuned model yields an additional 3.41 EM improvement on the open-source benchmark suite and 4.94 EM improvement on the private-domain benchmark suite. These findings indicate that \ourmethod{} is an effective augmentation to an optimized fine-tuned system.
\paragraph{Computation Cost}
While fine-tuning demonstrates substantial efficacy in enhancing code completion tasks, it is essential to consider the associated computational cost. Fine-tuning the \codellama{}-13B model requires substantial hardware resources, typically involving a cluster with eight NVIDIA A100 GPUs. In contrast, deploying \ourmethod{} is considerably more resource-efficient and operable on a single A100 GPU. In terms of computation time, fine-tuning requires a training duration of approximately 9.5 hours on the 8$\times$A100 cluster. Conversely, \ourmethod{} eliminates the need for training time, with the embedding and search processes aggregating to approximately 0.5 seconds on the same device.



%% file: tables/main_results_kwai.tex
\begin{table}[t]
    \centering
    \caption{Averaged results on private-domain benchmark}
    \begin{adjustbox}{width=0.8\columnwidth}
    \begin{tabular}{lcccc}
        \hline
        \multirow{2}{*}{\textbf{Method}} & \multicolumn{2}{c}{\textbf{\codellama{}}} & \multicolumn{2}{c}{\textbf{\starcoder{}}} \\
        \cline{2-5}
         & \textbf{EM} & \textbf{ES} & \textbf{EM} & \textbf{ES} \\
        \hline
        Base      & 33.67 & 63.30 & 32.03 & 62.14 \\
        \hline
        +BM25           & 41.84 & 67.02 & 40.76 & 65.67 \\
        +RepoCoder      & 41.70 & 66.95 & 40.62 & 65.60 \\
        +ReACC          & 41.63 & 66.87 & 40.59 & 65.53 \\
        \hline
        +Lexical Semantics   & 41.42 & 66.96 & 40.32 & 65.61 \\
        +Code Summary   & 42.10 & 67.24 & 41.07 & 66.85 \\
        +Hypo. Line     & 42.79 & 67.92 & 41.65 & 66.53 \\
        \hline
        +\ourmethod{}   & \textbf{46.07} & \textbf{71.13} & \textbf{44.82} & \textbf{68.62} \\
        \hline
    \end{tabular}
    \label{table:main_results_kwai}
    \end{adjustbox}
\end{table}


%% file: tables/prompting_template.tex
\begin{table}[h]
    \centering
    \caption{Prompting perspectives and templates. } 
    \begin{adjustbox}{width=1.0\columnwidth}
    \begin{tabular}{lcccc}
    \hline
    \textbf{No.} & \textbf{Perspect.} & \textbf{Prompt} & \textbf{EM} & \textbf{ES} \\
    \hline
    1 & Context & Embedding the following code snippets: [code]							& 50.44  & 73.83 \\
    2 & Context & Representing the following code snippets: [code]				        & 50.41  & 73.91 \\
    3 & Hypo. Line & [code]->[generation] 											    & 51.61  & 74.01 \\
    4 & Hypo. Line & Complete the code snippets [code]->[generation] 					& 51.21  & 74.18 \\
    5 & Summarization & This code snippets of [code] means -> [generation] 				    & 51.40  & 74.55 \\
    6 & Summarization & Summarize the code snippets [code] -> [generation] 			        & 51.08  & 74.02 \\
    \hline
    \end{tabular}
    \label{table:prompting_template}
    \end{adjustbox}
\end{table}

%% file: tables/retrieval_selection_retrievers.tex

\begin{table}[t]
\centering
\caption{Retrieval selection using different prompts}
\begin{adjustbox}{width=0.9\columnwidth}
\begin{tabular}{lccccc}
\hline
\textbf{Metrics} &\textbf{No.1+3} &\textbf{No.1+5} &\textbf{No.3+5} &\textbf{No.1+3+5} &\textbf{No.1-6} \\
\hline
EM & 52.43 & 52.94 & 53.51 & 54.66 & 54.88 \\
ES & 74.79 & 75.13 & 75.63 & 75.85 & 76.07 \\
\hline
\end{tabular}
\label{table:retrieval_selection_retriever}
\end{adjustbox}
\end{table}

%% file: tables/retrieval_selection.tex
\begin{table}[h]
\centering
\caption{Retrieval selection using different algorithms}
\begin{adjustbox}{width=0.55\columnwidth}
\begin{tabular}{lccc}
\hline
\textbf{Method} &\textbf{EM} & \textbf{ES} \\
\hline
Hypo. Line              &51.61   & 74.01\\
\hline
Union                   & 51.97  & 74.66 \\
Max Similarity 			& 52.37  & 74.68 \\
Logistic Regression 	& 53.21  & 75.28 \\
\hline
LinUCB                  & 54.66  & 75.85 \\
\hline
\end{tabular}
\label{table:retrieval_selection}
\end{adjustbox}
\end{table}

%% file: sections/6_threats.tex
\section{Threats to validity}
\paragraph{Internal Validity.}
The internal threat to validity lies in potential implementation bugs. To mitigate this, for compared techniques, we obtained original source code from GitHub repositories and used identical hyperparameters from their papers. And we have conducted a thorough review of our code scripts to ensure their correctness.

\paragraph{External Validity.}
The external threats to validity mainly lie in the benchmarks and techniques studied. To reduce these threats, we not only used established benchmarks but also included industry data unknown to LLMs. Through an exhaustive literature review, we believe the compared RAG-sequence models are sufficiently representative. Another threat is randomness in results. To alleviate this threat, we averaged results over five runs, reducing variance.

\paragraph{Construct Validity.}
The construct threat to validity lies in our evaluation metrics. Following previous work~\cite{zhang2023repocoder, lu2022reacc}, we adopted two widely-used metrics---\ExactMatch{} and Edit Similarity to comprehensively assess performance. Using established metrics provides rigorous quantification of improvements.

%% file: sections/7_related_work.tex
\section{Related Work}

\paragraph{Language Model for Code Completion}
To generate code completions of arbitrary lengths, researchers view code as a distinct variant of language and have subsequently used natural language processing techniques (NLP) to model code statistically. Earlier work leveraged N-gram models ~\cite{code_complete_Ngram}, recurrent neural networks such as LSTM ~\cite{code_complete_LSTM}, and attention mechanisms ~\cite{code_complete_attention} to encode programming languages. With the emergence of transformer-based models, language models (LMs) are trained on large-scale code datasets, which has significantly advanced code completion.
CodeBERT ~\cite{Feng2020CodeBERTAP}, one of the pioneering code LMs, performs the code completion task through masked language modeling. To facilitate the generation capability, later LMs mainly adopt either a decoder-only or an encoder-decoder model, which is trained to predict the subsequent token in an auto-regressive manner. For example, CodeGPT ~\cite{codexglue}, which follows the architecture of decoder-only GPT ~\cite{gpt2}, outperforms GPT2 in the code completion task. UniXCoder ~\cite{Guo2022UniXcoderUC}, a mixed encoder-decoder model, integrates multi-task learning strategies and leverages code structures to enhance pre-training and further advances code completion performance.
More recent LLMs, such as Codex ~\cite{chen2021evaluating}, CodeGen ~\cite{nijkamp2023codegen}, InCoder ~\cite{InCoder}, StarCoder ~\cite{starcoder}, and \codellama{} ~\cite{roziere2023codellama}, employ billions of parameters and trained on trillions of code tokens, significantly excel in code generation tasks. 
Notably, \codellama{} ~\cite{roziere2023codellama} adopts the fill-in-the-middle pretraining objective ~\cite{openaifim}, which resembles incomplete code contexts in code completion. This provides useful inductive bias, enabling Code Llama to substantially outperform prior models on infilling benchmarks ~\cite{InCoder}.

\paragraph{Retrieval Augmented Code Completion}
Retrieval augmented generation~\cite{meta2020rag} (RAG) has emerged as a technique to inject external knowledge into large language models (LLMs) to assist coherent text generation and mitigate hallucination for code completion. The RAG paradigm typically first retrieves the most relevant information using similarity measures such as BM25, dense embeddings such as SimCSE~\cite{Gao2021SimCSESC} or Dense Passage Retrieval~\cite{Karpukhin2020DensePR} (DPR). The retrieved information is then concatenated with the original input to guide the generation of LLM. Although initially explored for open-domain question answering, RAG has recently been adapted for code completion~\cite{lu2022reacc, zhang2023repocoder, knm}.
Early work in code completion ~\cite{lu2022reacc} focused on code-to-code retrieval using dual encoder models with the retrieved results fed to autoregressive LMs. While RepoCoder ~\cite{zhang2023repocoder} advances retrieval by iterating with incremental generations [7], KNM ~\cite{knm} incorporates in-domain code databases and utilizes Bayesian inference to finalize the code. 
Some other research works focus on cross-file retrieval, i.e., drawing context from cross-file context dependencies like imported libraries (e.g., ``$from \; transformers \; import \; GPTModelForCLM$'') or header files (``$include \; bta\_hh\_co.h$''). CrossCodeEval~\cite{Ding2023CrossCodeEvalAD} and RepoBench~\cite{Liu2023RepoBenchBR} construct benchmarks for such scenarios, while CocoMic~\cite{Ding2022CoCoMICCC} develops a cross-file context finder CCFINDER to identify and retrieve the most relevant cross-file context and integrates cross-file context to learn the in-file and cross-file context jointly atop pre-trained code LLMs.

In this paper, we propose \ourmethod{}, a code completion framework leveraging prompt engineering and contextual multi-armed bandit for the first time to flexibly incorporate and adapt to multiple perspectives of code. Our extensive evaluation results indicate that \ourmethod{} can significantly enhance the code completion effectiveness over the existing RAG-based code completion techniques, indicating the strengths of our proposed RAG mechanisms.

%% file: sections/8_conclusion.tex
\section{Conclusion}
In this paper, we propose \ourmethod{}, the first code completion technique to integrate prompt engineering and contextual multi-armed bandit to flexibly incorporate and adapt to multiple perspectives of code. \ourmethod{} first employs a \PromptRetrieval{} which crafts prompt templates to elicit LLM knowledge to understand code semantics with multiple retrieval perspectives. Then, it adopts the \PromptBandit{} to incorporate code similarity into the decision-making process to determine the most suitable retrieval perspective for the LLM to complete the code. 
Extensive evaluations across both open-source and private-domain repositories demonstrate the superior performance and adaptability of \ourmethod{}, marking a significant advancement over the state-of-the-art technique by 8.6\% and 10.1\% in terms of EM respectively.

\section{Data Availability}
Considering the deployment of the artifact within the e-commerce company and the privacy protection policy, the dataset containing the private-domain code of the company shall remain undisclosed. Nonetheless, we provide the repository~\cite{procc} for all the other available materials, including the source code of the artifact and the open-source dataset.